%% file: main_cr_with_appendix.tex
\documentclass[letterpaper]{article} 
\usepackage{aaai24}  
\usepackage{times}  
\usepackage{helvet}  
\usepackage{courier}  
\usepackage[hyphens]{url}  
\usepackage{graphicx} 
\usepackage{natbib}  
\usepackage{caption} 
\usepackage{algorithm}
\usepackage{algorithmic}
\usepackage{newfloat}
\usepackage{listings}
\DeclareCaptionStyle{ruled}{labelfont=normalfont,labelsep=colon,strut=off} 
\floatstyle{ruled}
\newfloat{listing}{tb}{lst}{}
\floatname{listing}{Listing}
\input{macros_danceanyway}

\title{\methodname: Synthesizing Beat-Guided 3D Dances with Randomized Temporal Contrastive Learning}
\author{
    Aneesh Bhattacharya\textsuperscript{\rm 1,2},
    Manas Paranjape\textsuperscript{\rm 1},
    Uttaran Bhattacharya\textsuperscript{\rm 3},
    Aniket Bera\textsuperscript{\rm 1}
}
\affiliations{
    \textsuperscript{\rm 1}Purdue University, USA \\
    \textsuperscript{\rm 2}IIIT Naya Raipur, India \\
    \textsuperscript{\rm 3}Adobe Research, USA \\

    \{bhatta95, mparanja, aniketbera\}@purdue.edu, ubhattac@adobe.com
}

\begin{document}

\maketitle

\input{abstract}

\section{Introduction}\label{sec:intro}

Dancing is a central human behavior observed across societies and cultures~\cite{dancing_species}. Being simultaneously a form of expression and communication, the space of dance motions is dense, diverse, and, at the same time, temporally cohesive and structured~\cite{edge}. The complexity of dance motions and their pervasiveness in our socio-cultural fabric has led to extensive research on generating dancing digital characters for applications such as character design~\cite{game_development}, storyboard visualization for consumer media~\cite{gesticulator,script_visualization}, building metaverse tools~\cite{omniverse} and even advancing our understanding of the relationships between music and dance~\cite{neuroscience_of_dance}.

\begin{figure*}[t]
    \centering
    \includegraphics[width=0.9\textwidth]{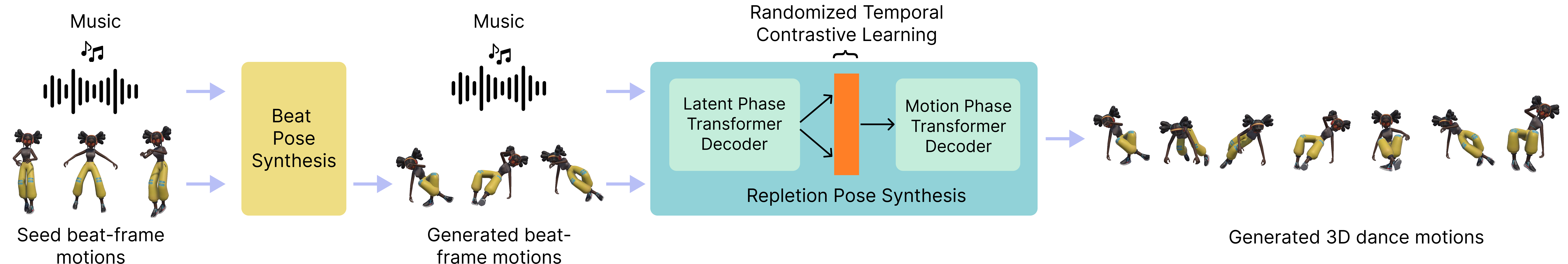}
    \caption{\methodname: A two-stage hierarchical network that can generate beat-aligned and diverse, fine-grained 3D dances given audio. We render our results with Mixamo characters.}
    \label{fig:teaser}
\end{figure*}

Prior methods in dance generation can adapt to different dance genres but may encounter temporal inconsistencies~\cite{graph_dance1} and motion freezing and instability~\cite{ai_choreo}. Using seed poses is a common approach to enforce plausibility~\cite{ai_choreo,dancenet,transformer_dance1}. However, they only provide the initial dance characteristics and become less relevant over time when generating long dance sequences. Diffusion-based approaches~\cite{edge} offer better control in the generative process but come at the cost of slow inference speed and heavy parameter tuning for novel datasets. Other approaches tokenize the dance sequences into a finite, learnable set of quantized vectors~\cite{bailando}, which can generate long sequences with minimal tuning but trade-off on the fine-grained diversity of the generated dances.

Different from these approaches, we make two key observations. First, dancers often exhibit bursts or drops of energy at the audio beats. Therefore, the dance steps at the audio beats provide a coarse structure of the dance. Second, we can forego tokenization and explicitly enforce temporal diversity in the generative process to get long dance sequences in the continuous space without freezing or collapsing to repetitive patterns. In this paper, we introduce our method \textit{\methodname~} (Fig.~\ref{fig:teaser}), built on these observations, to generate plausible 3D dances from audio. We learn the correlation between the dance motions and the audio at two temporal levels: a coarser \textit{beat level}, which corresponds to the dance poses at the audio beat frames, and a finer \textit{repletion level}, which corresponds to the dance poses at all the other frames. Further, we perform a randomized temporal contrastive loss between segments at the repletion level to enforce diversity of motion between segments that are arbitrarily far from each other. By explicitly learning the correlation between the audio and the dance poses at the beat frames, we can generate plausible beat pose sequences representing the underlying dance characteristics for the entirety of the audio. Given these beat poses, we can generate the remaining or repletion poses while ensuring they are sufficiently diverse.

In summary, our main contributions are as follows:
\begin{itemize}
    \item A temporally hierarchical learning method using sequence-to-sequence and generative adversarial learning to synthesize beat-aligned 3D dance sequences for digital characters synchronized with audio.
    \item A randomized temporal contrastive loss to generate fine-grained, diverse motions, particularly in the long term.
    \item Leveraging the spatial-temporal graph representation of the 3D human poses to efficiently learn both the localized (joint-level) and the macroscopic (body-level) movements for different dances.
    \item An end-to-end pipeline for audio-to-dance generation, which exhibits \sota~performance on multiple quantitative and qualitative evaluations.
\end{itemize}

\begin{figure*}[t]
    \centering
    \includegraphics[width=0.86\textwidth]{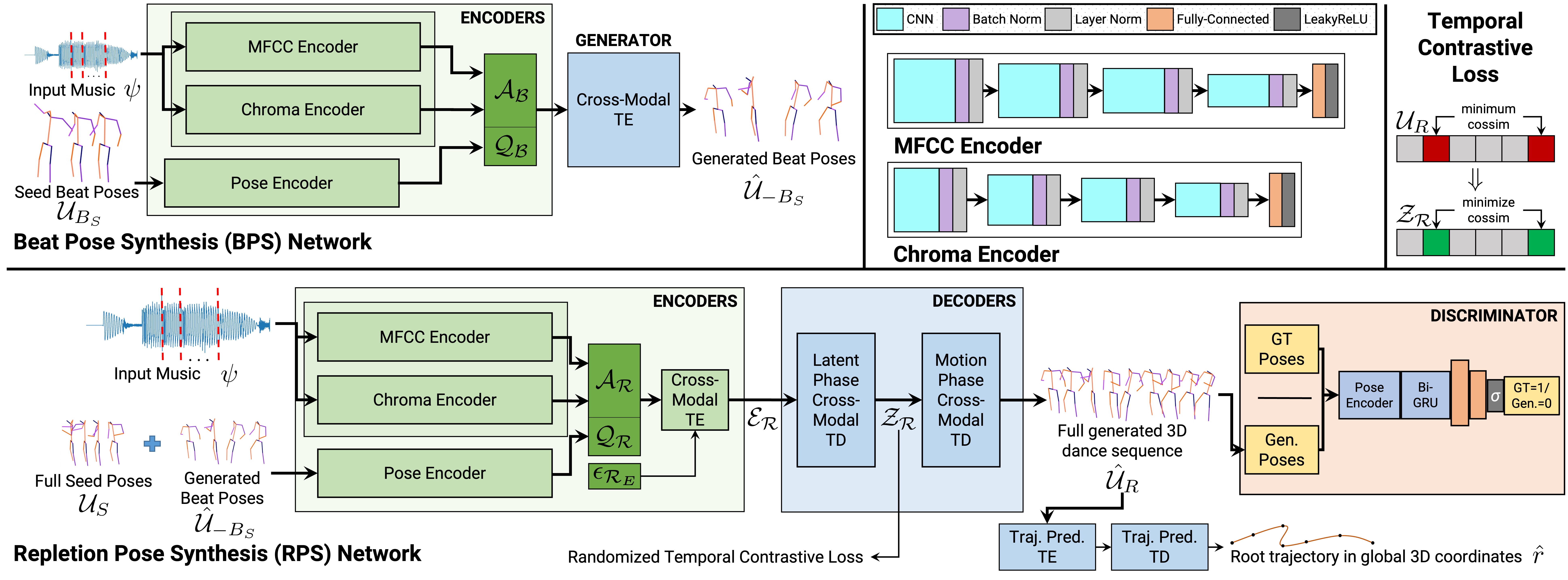}
    \caption{\methodname~Network Architecture. \methodname~consists of two stages, Beat Pose Synthesis (BPS) and Repletion Pose Synthesis (RPS), trained one after the other. BPS \textit{(top row, left)} has a predictor architecture to generate the coarse beat poses, and RPS \textit{(bottom row)} has a generative adversarial architecture to generate all the remaining poses with fine-grained detail, followed by a seq-to-seq trajectory predictor for the global root translations. To train our RPS, we propose an additional randomized temporal contrastive loss \textit{(top row, right)} to enforce motion diversity. For completeness, we also expand our MFCC and Chroma encoders \textit{(top row, middle)}, which have the same architecture but different layer sizes.}
    \label{fig:dance_synth_net}
\end{figure*}


\section{Related Work}\label{sec:rw}
We briefly review methods for human motion synthesis, particularly from audio inputs such as speech and music.

\paragraph{3D Human Motion-to-Motion Synthesis.}
3D motion-to-motion synthesis is richly explored in computer vision and graphics. Classical approaches include kernel-based probability distributions~\cite{kernel3,kernel4} to predict the most likely future poses given past poses, and motion graphs~\cite{Arikan2002InteractiveMG,Lucas} to represent poses as nodes in a graph and transitioning between those poses according to various linking rules. These models often require significant manual tuning and do not allow for the incorporation of additional input modalities.
More recently, learning-based approaches have gained immense traction in this area through convolutional networks~\cite{cnn1,cnn2}, recurrent networks~\cite{rnn1,rnn2,rnn3,rnn4,rnn5,rnn6,rnn7,rnn8,npss}, generative adversarial networks~\cite{gan1}, graph convolutional networks~\cite{gcn1}, and transformers \cite{transf1,transf2}. Current methods achieve high-quality performance on large-scale datasets and can generate diverse and realistic motions. 
However, these learning-based approaches are autoregressive and do not condition the motions on additional modalities such as audio.

\paragraph{3D Dance Motion Synthesis from Audio.}
Procedural methods for audio-to-dance synthesis use approaches such as motion graphs, where the audio rhythms are used to constrain the graph linking rules~\cite{graph_dance1,graph_dance2, pan2021diverse, yang2023keyframe,aristidou2021rhythm, chen2021choreomaster}. However, these approaches may suffer from temporal conflicts due to the differences in dance tempos. More recent approaches generate 3D dance motions using deep neural networks.
LSTM-based methods~\cite{lstm_dance1,lstm_dance2} can synthesize long, complex dance sequences by modeling the temporal dependencies in the motion data. GAN-based methods~\cite{gan_dance1,graph_dance2} train a generator network to produce realistic dance sequences that match the distribution of a given dataset and a discriminator network to distinguish between the generated and real dance sequences.
Transformers-based methods leverage self- and cross-attention mechanisms to capture long-range dependencies between the audio and the dances~\cite{transformer_dance1,ai_choreo,edge}. To overcome the transformers' limitations in generating continuous-space sequences, some transformer variants condense the latent space of dances into a finite set of quantized vectors~\cite{bailando}. Other methods segregate the learning into two steps: first generating key poses and then interpolating between them~\cite{danceformer}. They have also been paired with diffusion models~\cite{edge} to enhance joint-level editing and motion in-betweening capabilities.
Large-scale 3D MoCap dance datasets~\cite{groovenet,lstm_dance1,dancenet} have played a crucial role in the success of these methods. These datasets have been used to train and test the generative models. Additionally, 3D human models have been mapped to the motion data~\cite{aist++}, leading to the generation of realistic and expressive dance motions.
However, these methods can sometimes result in non-standard poses or regression to mean configurations without exhibiting animated movements due to the high dimensionality of long pose sequences, or cannot adapt to fine-grained dance motions due to quantization. For interpolation-based methods, any error in key pose generation gets propagated to the interpolation network during inference. To overcome these limitations, our method explicitly learns the beat poses to ensure long-term beat alignment and performs a randomized temporal contrastive loss between segments to ensure fine-grained diversity of motions. We also use our generated beat poses only as control signals for interpolation, as a result of which the interpolation process can generate spatially and temporally plausible movements regardless of errors in the beat pose generation.

\paragraph{3D Human Motion Synthesis from Other Modalities.}
Besides music, 3D motions are commonly generated using other modalities, such as speech and text. Co-speech gesture synthesis methods generate accompanying gestures for speech based on learned individual gesticulation patterns. These approaches aim to personalize the synthesis process by capturing the individual characteristics of the speaker~\cite{individual_gesture_styles}, enhance generative capabilities using GAN-based approaches~\cite{multi_adversarial_gestures}, enhance robustness by combining speech, text and speaker identities in the inputs~\cite{trimodal}, incorporate emotional cues from the audio and gesticulation patterns~\cite{s2ag}, explicitly add rhythm-aware information~\cite{rhythmic_gesticulator}, and use diffusion models to enable editability~\cite{gesturediffuclip}.
In our work, we leverage the audio beat information and the physiological dance movements and use an adversarial framework to improve the plausibility of the generated dances.


\section{Temporally Hierarchical Dance Synthesis}\label{sec:approach}
We aim to generate 3D pose sequences for dances given input audio. Our approach is to separately learn the \textit{structure} of the dance described by the \textit{beat poses} or the poses at the beat frames of the audio, and the \textit{finer details} of the dance described by the \textit{repletion poses} or the poses at the remaining frames. To this end, we develop a two-stage learning method consisting of Beat Pose Synthesis (BPS) followed by Repletion Pose Synthesis (RPS). In BPS, given a short sequence of seed beat poses and the audio, we generate the beat poses. In RPS, given all the seed poses, the beat poses following the seed pose duration, and the audio, we generate the remaining poses to complete the dance. Mathematically, we represent the pose at frame $t$ as $\mathcal{U}_t = \bracks{u_t^{\parens{1}}, \dots, u_t^{\parens{J-1}}} \in \mathbb{R}^{\parens{J-1} \times 3}$, consisting of the unit line vectors denoting the $J-1$ bones corresponding to the $J$ body joints. We take in the audio as a raw waveform and process it into a feature sequence $\mathcal{A} = \bracks{a_1, \dots, a_T} \in \mathbb{R}^{D_\mathcal{A} \times T}$ for some feature dimension $D_\mathcal{A}$ and total temporal length $T$. We extract the beat frames from the audio using available beat detection methods and represent them as a set $B = \braces{\textrm{beat frames in }\mathcal{A}}$. These beat frames may or may not be equidistant in time. Our BPS takes in the audio features $\mathcal{A}$ and the initial seed beat pose sequence $\mathcal{U}_{B_S} = \braces{\mathcal{U}_f}_{f \in {B_S}}$, where $B_S \subset B$ consists of all the beat frames in $B$ contained within the seed sequence length $T_S \ll T$, and generates the beat poses corresponding to $\mathcal{U}_{-B_S} = \braces{\mathcal{U}_f}_{f \in B - B_S}$. Our RPS takes in the audio features $\mathcal{A}$, all the seed poses $\mathcal{U}_S = \braces{\mathcal{U}_f}_{f \in \braces{1, \dots, T_S}}$, and the generated beat poses corresponding to $\mathcal{U}_{-B_S} = \braces{\mathcal{U}_f}_{f \in {-B_S}}$, and synthesizes the repletion poses corresponding to $\mathcal{U}_R = \braces{\mathcal{U}_f}_{f \in R}$ where $R = \braces{1, \dots, T} - \parens{B \cup S}$. We first fully train our BPS and then use its generated outputs to fully train our RPS, followed by a trajectory predictor for the global root translations. We show the overview of our end-to-end pipeline in Fig.~\ref{fig:dance_synth_net} and describe the individual components below.

\subsection{Beat Pose Synthesis}\label{subsec:bps}
Our Beat Pose Synthesis (BPS) network takes in the raw audio waveform $\psi$ and the seed beat pose sequence $\mathcal{U}_{B_S}$, and generates the beat poses $\hat{\mathcal{U}}_{-B_S}$. It uses multiple feature encoders to extract rhythmic and semantic information from the audio and the physiological information from the seed beat poses. It combines these features through a transformer-encoder-based generator to synthesize the beat poses.

\paragraph{Feature Encoders.}
We encode the audio and the seed pose sequences using separate encoder blocks. The \textit{audio encoder block} consists of an MFCC encoder and a Chroma encoder. MFCCs naturally capture the human auditory response and are commonly used in tasks such as emotion recognition~\cite{mfcc_ser} and speaker identification~\cite{mfcc_speaker_identification}. In our work, we leverage the audio prosody and the vocal intonations (when present) captured by the MFCCs. An MFCC encoder $\mathcal{M}_{\mathcal{B}}$ takes in the MFCCs and their first- and second-order derivatives and uses convolutional layers to learn $D_M$-dimensional latent feature sequences $\mathcal{A}_{M_{\mathcal{B}}} \in \mathbb{R}^{D_M \times \modulus{B}}$ from their localized inter-dependencies as
\begin{equation}
    \mathcal{A}_{M_{\mathcal{B}}} = \mathcal{M}_{\mathcal{B}}\parens{\textrm{MFCC}\parens{\psi}; W_{M_{\mathcal{B}}}},
\end{equation}
where $W_{M_{\mathcal{B}}}$ are the trainable parameters. Chroma CENS features capture the melody and pitch in audio, and we use a Chroma encoder $\mathcal{C}_{\mathcal{B}}$ with convolutional layers to transform these Chroma CENS features into $D_{C_{\mathcal{B}}}$-dimensional latent feature sequences $\mathcal{A}_{C_{\mathcal{B}}} \in \mathbb{R}^{D_{C_{\mathcal{B}}} \times \modulus{B}}$ based on their localized inter-dependencies as
\begin{equation}
    \mathcal{A}_{C_{\mathcal{B}}} = \mathcal{C}_{\mathcal{B}}\parens{\textrm{Chroma}\parens{\psi}; W_{C_{\mathcal{B}}}},
\end{equation}
where $W_{C_{\mathcal{B}}}$ are the trainable parameters. We concatenate these two features into audio features $\mathcal{A}_{\mathcal{B}}$ as
\begin{equation}
    \mathcal{A}_{\mathcal{B}} = \bracks{\mathcal{A}_{M_{\mathcal{B}}}; \mathcal{A}_{C_{\mathcal{B}}}} \in \mathbb{R}^{D_{\mathcal{A}_{\mathcal{B}}} \times \modulus{B}},
\end{equation}
where $D_{\mathcal{A}_{\mathcal{B}}} = D_M + D_{C_{\mathcal{B}}}$. For the \textit{pose encoder block}, we adopt the pose encoder architecture of~\cite{s2ag} to learn the physiological variations in the dance motions represented by $\mathcal{U}_{B_S}$. The pose encoder block $\mathcal{P}_{\mathcal{B}}$ outputs latent pose features $\mathcal{Q} \in \mathbb{R}^{D_P \times \modulus{B}}$ as
\begin{equation}
    \mathcal{Q}_{\mathcal{B}} = \mathcal{P}_{\mathcal{B}}\parens{\mathcal{U}_{B_S}; W_{P_{\mathcal{B}}}},
\end{equation}
where $W_{P_{\mathcal{B}}}$ are the trainable parameters.

\paragraph{Transformer-Encoder-Based Generator.}
We concatenate the latent features $\mathcal{A}_{\mathcal{B}}$ and $\mathcal{Q}_{\mathcal{B}}$ and pass them through a transformer encoder (TE) $\Theta_{\mathcal{B}}$ with cross-attention between the two features to generate $\hat{\mathcal{U}}_{-B_S}$, as
\begin{equation}
    \hat{\mathcal{U}}_{-B_S} = \Theta_{\mathcal{B}}\parens{\mathcal{A}_{\mathcal{B}} \oplus \mathcal{Q}_{\mathcal{B}}; W_{\Theta_{\mathcal{B}}}},
\end{equation}
where $\oplus$ denotes concatenation and $W_{\Theta_{\mathcal{B}}}$ are the trainable parameters. We note the use of TE here, which generates the entire sequence at once. This is because the beat frames can be irregularly separated in time, and the traditional autoregressive decoder fails to learn these separations.

\begin{figure}
    \centering
    \includegraphics[width=0.75\columnwidth]{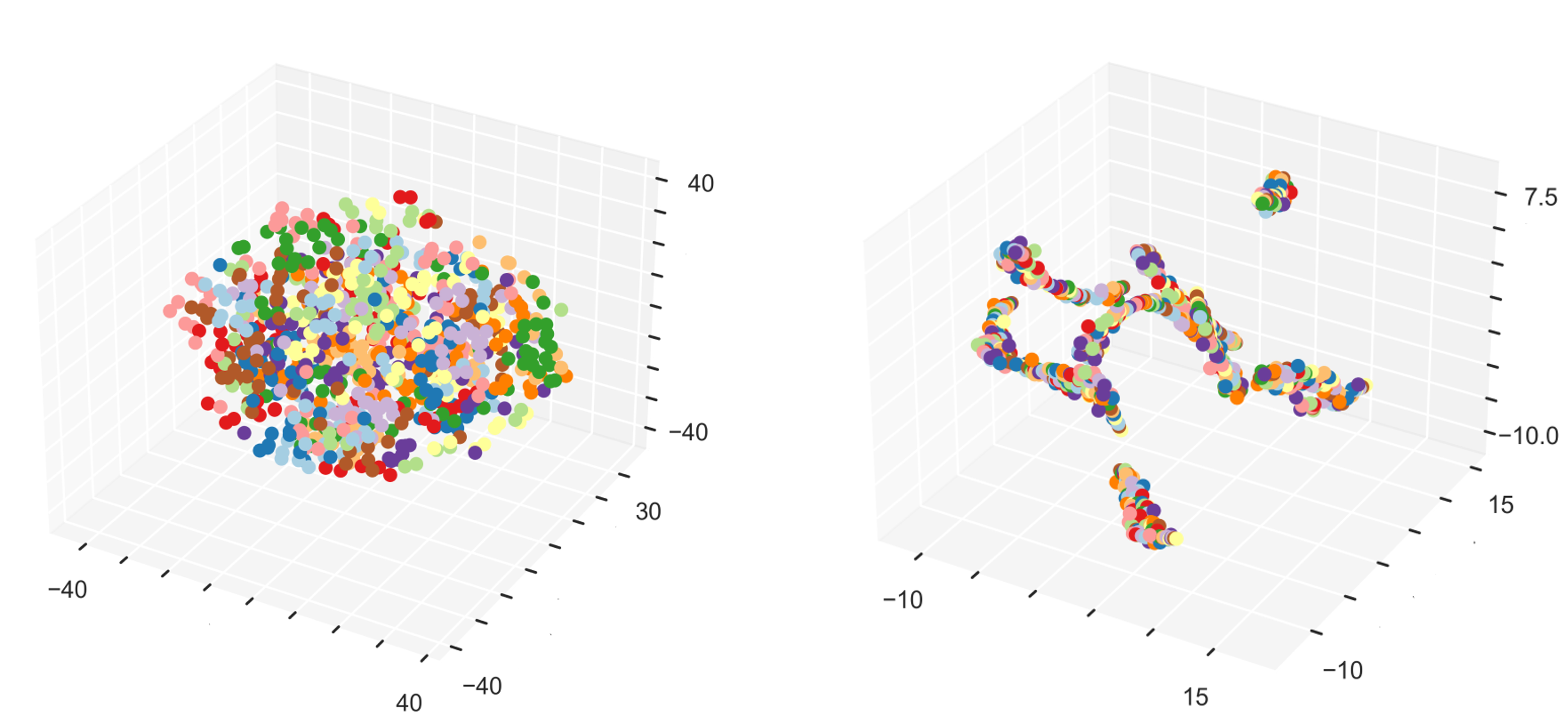}
    \caption{$t$-SNE Plot of Samples from RPS Latent Decoder Space. Distribution of the features for the $m$-length segments in $\mathcal{Z}_{\mathcal{R}}$, for 100 random samples (each represented with a different color) in AIST++~\cite{aist++}, after training with \textit{(right)} and without \textit{(left)} our RTC loss. Clustering all the sample segments using the RTC loss is necessary to generate diverse motions.}
    \label{fig:rtc_tsne}
\end{figure}

\subsection{Repletion Pose Synthesis}
In contrast to BPS, our Repletion Pose Synthesis (RPS) network generates a dense sequence of repletion poses, capturing the finer details. Traditional seq-to-seq approaches fail to capture these details and lead to mean regression. To overcome this, we opt for a generative adversarial approach using a generator and a discriminator. The generator takes in the raw audio waveform $\psi$, the \textit{full} seed pose sequence $\mathcal{U}_{S}$, all the generated beat poses $\hat{\mathcal{U}}_{-B_S}$ and Gaussian noise $\epsilon_{\mathcal{R}_E}$ for the encoder, and synthesizes the repletion poses $\hat{\mathcal{U}}_R$. The discriminator learns to distinguish between the ground truth and the synthesized pose sequences based on their physiological features, and the generator eventually produces dance motions that the discriminator cannot distinguish from the ground truth, leading to plausible synthesized dances.

\paragraph{Generator.}\label{para:generator}
The RPS generator consists of encoder blocks (similar to the BPS network), followed by a transformer encoder-decoder (TE-TD) architecture with cross-attention. Specifically, we use an MFCC encoder $\mathcal{M}_{\mathcal{R}}$, a Chroma encoder $\mathcal{C}_{\mathcal{R}}$, and a pose encoder $\mathcal{P}_{\mathcal{R}}$, similar in architecture to their BPS counterparts but trained separately with parameters $W_{\mathcal{M}_{\mathcal{R}}}$, $W_{\mathcal{C}_{\mathcal{R}}}$, and $W_{\mathcal{P}_{\mathcal{R}}}$, to obtain the counterpart latent features $\mathcal{A}_{\mathcal{R}}$ and $\mathcal{Q}_{\mathcal{R}}$. Different from BPS, we include $\epsilon_{\mathcal{R}_E}$ as an additional input feature and then use the TE $\Theta_{\mathcal{R}}$ to obtain encoded features $\mathcal{E}_{\mathcal{R}}$, as 
\begin{equation}
    \mathcal{E}_{\mathcal{R}} = \Theta_{\mathcal{R}}\parens{\mathcal{A}_{\mathcal{R}} \oplus \mathcal{Q}_{\mathcal{R}} \oplus \epsilon_{\mathcal{R}_E}; W_{\Theta_{\mathcal{R}}}},
\end{equation}
where $W_{\Theta_{\mathcal{R}}}$ are the trainable parameters. We decode $\mathcal{E}_{\mathcal{R}}$ first into a latent space and then into the motion space, both employing transformer decoders (TDs), as
\begin{equation}
    \mathcal{Z}_{\mathcal{R}} = \Phi_{\mathcal{R}}^Z\parens{\mathcal{E}_{\mathcal{R}}; W_{\Phi_{\mathcal{R}}^Z}},
    \label{eq:RPS_latent_decoder}
\end{equation}
where $\mathcal{Z}_R = \braces{\mathcal{Z}_f  \in \mathbb{R}^{D_Z}}_{f \in R}$ are the $D_Z$-dimensional latent features and $W_{\Phi_{\mathcal{R}}^Z}$ are the trainable parameters. In the subsequent motion decoding phase, we use a TD $\Phi_{\mathcal{R}}^U$, as
\begin{equation}
    \hat{\mathcal{U}}_{\mathcal{R}} = \Phi_{\mathcal{R}}^U\parens{\mathcal{Z}_{\mathcal{R}}; W_{\Phi_{\mathcal{R}}^U}},
\end{equation}
where $W_{\Phi_{\mathcal{R}}^U}$ are the trainable parameters. We perform latent decoding into $\mathcal{Z}_R$, of the same sequence length as $\hat{\mathcal{U}}_{\mathcal{R}}$, to efficiently apply temporal diversity constraints on the smooth, equivalence space of $\mathcal{Z}_R$ rather than on the non-smooth space of unit line vector sequences $\hat{\mathcal{U}}_{\mathcal{R}}$.

\paragraph{Discriminator.}\label{para:discriminator}
Our discriminator takes in 3D dance pose sequences $\widetilde{\mathcal{U}}_{\mathcal{R}}$, which can be either the ground truth $\mathcal{U}_{\mathcal{R}}$ or the generated $\hat{\mathcal{U}}_{\mathcal{R}}$, and uses a pose encoder $\mathcal{P}_{\mathcal{D}}$ (same architecture as $\mathcal{P}_{\mathcal{R}}$) to learn latent pose features $\widetilde{\mathcal{Q}}_{\mathcal{D}} \in \mathbb{R}^{D_P \times \modulus{R}}$ based on the physiological variations in the dances, as
\begin{equation}
    \widetilde{\mathcal{Q}}_{\mathcal{D}} = \mathcal{P}_{\mathcal{D}}\parens{\widetilde{\mathcal{U}}; W_{P_{\mathcal{D}}}},
\end{equation}
where $W_{P_{\mathcal{D}}}$ are the trainable parameters. It then uses a bidirectional GRU (BiGRU) of latent dimension $H_{\mathcal{D}}$ to learn the temporal inter-dependencies in the pose features, followed by a set of FC layers to compress the features into scalar variables and a sigmoid function to compute binary class probabilities $p_{\mathcal{D}} \in \bracks{0, 1}$, as
\begin{equation}
    p_{\mathcal{D}} = \sigma\parens{\textrm{FC}_{\mathcal{D}}\parens{\textrm{BiGRU}_{\mathcal{D}}\parens{\widetilde{\mathcal{Q}}_{\mathcal{D}}; W_{L_{\mathcal{D}}}}; W_{FC_{\mathcal{D}}}}},
\end{equation}
where we only consider the output of the BiGRU and not its hidden states, $W_{L_{\mathcal{D}}}$ and $W_{FC_{\mathcal{D}}}$ denote the trainable parameters, and $\sigma\parens{\cdot}$ denotes the sigmoid function.

\paragraph{Trajectory Predictor.}\label{para:translation_decoder}
We complete the dances by learning the root trajectory from the generated poses. We use a TE-TD architecture that takes in $\hat{\mathcal{U}}_{\mathcal{R}}$, learns latent pose features $\mathcal{Q}_{\mathcal{T}} \in \mathbb{R}^{D_P \times \modulus{R}}$ through a TE $\Theta_{\mathcal{T}}$ with trainable parameters $W_{\Theta_{\mathcal{T}}}$, and decodes them autoregressively through a TD $\Phi_{\mathcal{T}}$ with trainable parameters $W_{\Phi_{\mathcal{T}}}$ to predict the 3D world coordinates of the root, $\hat{r} \in \mathbb{R}^{3 \times \modulus{R}}$.

\begin{table}[t]
    \centering
    \resizebox{\columnwidth}{!}{
        \begin{tabular}{lrrrrrr}
            \toprule
            Method & \multicolumn{2}{c}{Quality} & \multicolumn{2}{c}{Diversity} & BAS & PFC \\
            \cmidrule(lr){2-3}\cmidrule(lr){4-5}
            & $\textrm{FID}_k$ & $\textrm{FID}_g$ & $\textrm{MD}_k$ & $\textrm{MD}_g$ && \\
            & $\downarrow$ & $\downarrow$ & $\rightarrow$ & $\rightarrow$ & $\uparrow$ & $\downarrow$ \\
            \midrule
            GT & 17.10 & 10.60 & 10.61 & 7.48 & 0.24 & 0.32 \\
            \midrule
            Dnc. Tf. & 86.43 & 43.46 & 6.85 & 3.32 & 0.16 & $\times$ \\
            DncNet & 69.18 & 25.49 & 2.86 & 2.85 & 0.14 & $\times$ \\
            DncRev. & 73.42 & 25.92 & 3.52 & 4.87 & 0.19 & $\times$ \\
            FACT & 35.35 & 22.11 & 10.85 & 6.14 & 0.22 & 2.25 \\
            B'lando & 28.16 & 9.62 & 7.92 & 7.72 & 0.23 & 1.75 \\
            EDGE & \underline{20.55} & \underline{9.49} & \underline{10.58} & \underline{7.62} & \underline{0.27} & \underline{1.65} \\
            \midrule
            \textbf{\methodnameshort} & \textbf{17.98} & \textbf{9.42} & \textbf{10.62} & \textbf{7.51} & \textbf{0.33} & \textbf{0.83} \\
            $\,-$ BPS & 18.64 & 9.73 & 7.26 & 4.98 & 0.22 & 1.79 \\
            $\,-$ RTC & 18.54 & 9.95 & 6.91 & 4.81 & 0.26 & 1.05 \\
            $\,-$ rand. & 18.82 & 11.08 & 7.72 & 5.28 & 0.27 & 2.80 \\
            \bottomrule
        \end{tabular}
    }
    \caption{Quantitative Evaluation on AIST++~\cite{aist++}. Bold = best, underline = best among current methods, $\times$ = metric not calculated, arrows are directions for better values: $\uparrow$ = higher, $\downarrow$ = lower, $\rightarrow$ = closer to ground truth.}
    \label{tab:quantitative_eval_aist}
\end{table}


\section{Training and Testing}
We detail the loss functions we use for training our network, the implementation details, and the testing procedure.

\subsection{Training Loss Functions}\label{subsec:training_losses}
We use pose and leg motion losses to first train our BPS. We then use our proposed randomized temporal contrastive (RTC) loss, pose and leg motion losses, and adversarial losses to train our RPS. We describe our RTC loss below and provide details of the other losses in our appendix.

\paragraph{Randomized Temporal Contrastive (RTC) Loss.}
While the transformer is currently \sota~for sequence generation~\cite{attention_is_all_you_need}, it can lead to freezing and mode collapse for motion sequences such as dances~\cite{aist++}, where the sequence variables lie in an infinite, continuous space rather than in a finite set of quantized tokens. To address these issues, we consider overlapping segments of length $m$ within each sequence with a sliding length $d$, and enforce diversity across these segments. We choose a segment $n$ at random and obtain the non-overlapping segment $\bar{n}$ with the minimum cosine similarity to it, as
\begin{equation}
    \bar{n} = \arg\min_{x \in N}\modulus{\textrm{cossim}\parens{\mathcal{U}_n, \mathcal{U}_x}},
\end{equation}
where $N$ is the set of $\ceil{\frac{\modulus{R} - m}{d}}$ segments. We then compute our RTC loss on the RPS latent decoder space (Eqn.~\ref{eq:RPS_latent_decoder}) as
\begin{equation}
    \mathcal{L}_{RTC} = \modulus{\textrm{cossim}\parens{\mathcal{Z}_n, \mathcal{Z}_{\bar{n}}}}.
\end{equation}
This ensures that the segments of our generated sequences are as temporally well-separated as the corresponding training data, enforcing diversity and avoiding freezing or collapse to repetitive motions. The random choice of $n$ is necessary as it prevents the network from memorizing segment positions and makes it focus on \textit{all} the segments across the training epochs in an expected sense. Using the smooth space of the latent decoder sequence $\mathcal{Z}$ instead of the non-smooth motion space of $\hat{\mathcal{U}}$ is also necessary, as it enables stable backpropagations.
Our RTC loss thus enables the transformer architecture to operate reliably on continuous sequences. This differs from the commonly used alternative of vector quantization (VQ) followed by tokenized sequence generation~\cite{vqvae}, which limits the generative power to the finite set of quantized vectors. While some methods improve on the conventional VQ approach by learning separate upper- and lower-body representations~\cite{bailando}, their combined representations cannot encompass all possible motions in the full-body motion space.

\begin{figure}[t]
    \centering
    \includegraphics[width=0.9\columnwidth]{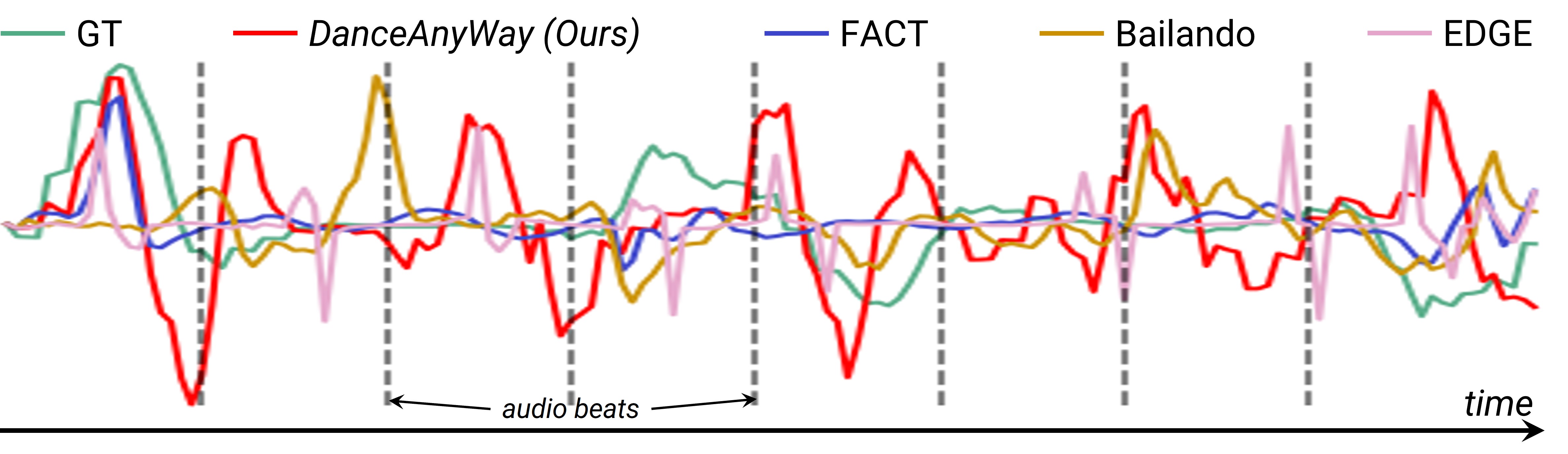}
    \caption{Beat Alignment. Kinetic velocities over time for one ground truth (GT) motion and corresponding generative results. Our method has more peaks and valleys at the beat frames, indicating more alignment with the audio.}
    \label{fig:beat_align}
\end{figure}

\subsection{Implementation Details}
We train our network using $7$-second dance clips sampled at $10$ fps, \textit{i.e.}, with $T=70$, and use a seed pose length $T_S = 20$. For our RTC loss, we use segments of length $m=25$ with sliding window length $d=5$. We use Librosa~\cite{librosa} to extract the MFCC and the Chroma CENS features and compute the beat frames. We use a maximum of $\modulus{B} = 20$ beat frames and $\modulus{B_S} = 3$ seed beat frames. We use $D_M=32$, $D_{C_{\mathcal{B}}}=4$, $D_{C_{\mathcal{R}}}=6$, $D_P=16$, $\Theta_{\mathcal{B}}$, $\Theta_{\mathcal{R}}$, and $\Theta_{\mathcal{T}}$. $\Theta_{\mathcal{B}}$ and $\Theta_{\mathcal{T}}$ have $4$ heads and $6$ blocks, while $\Theta_{\mathcal{R}}$ has $8$ heads and $6$ blocks, $\Phi_{\mathcal{R}}^Z$ with $3$ heads and $8$ blocks, $\Phi_{\mathcal{R}}^U$ with $3$ heads and $4$ blocks, and $\Phi_{\mathcal{T}}$ with $1$ head and $8$ blocks. For BPS, we use the Adam optimizer~\cite{adam} with $\beta_1=0.5$, $\beta_2=0.99$, a mini-batch size of $8$, a learning rate (LR) of $1e-4$, and train for $500$ epochs. For RPS, we use the Adam optimizer with $\beta_1=0.5$, $\beta_2=0.99$, a mini-batch size of $8$, an LR of $1e-4$ for both the generator and discriminator, and train for $250$ epochs. For our trajectory predictor, we use the Adam optimizer with $\beta_1=0.8$, $\beta_2=0.99$, a mini-batch size of $8$, a learning rate of $1e-5$, and train for a total of $700$ epochs.
Training our BPS, RPS, and trajectory predictor takes $6$, $16$, and $3$ hours, respectively, on an NVIDIA A100 GPU.

\begin{figure}[t]
    \centering
    \includegraphics[width=0.84\columnwidth]{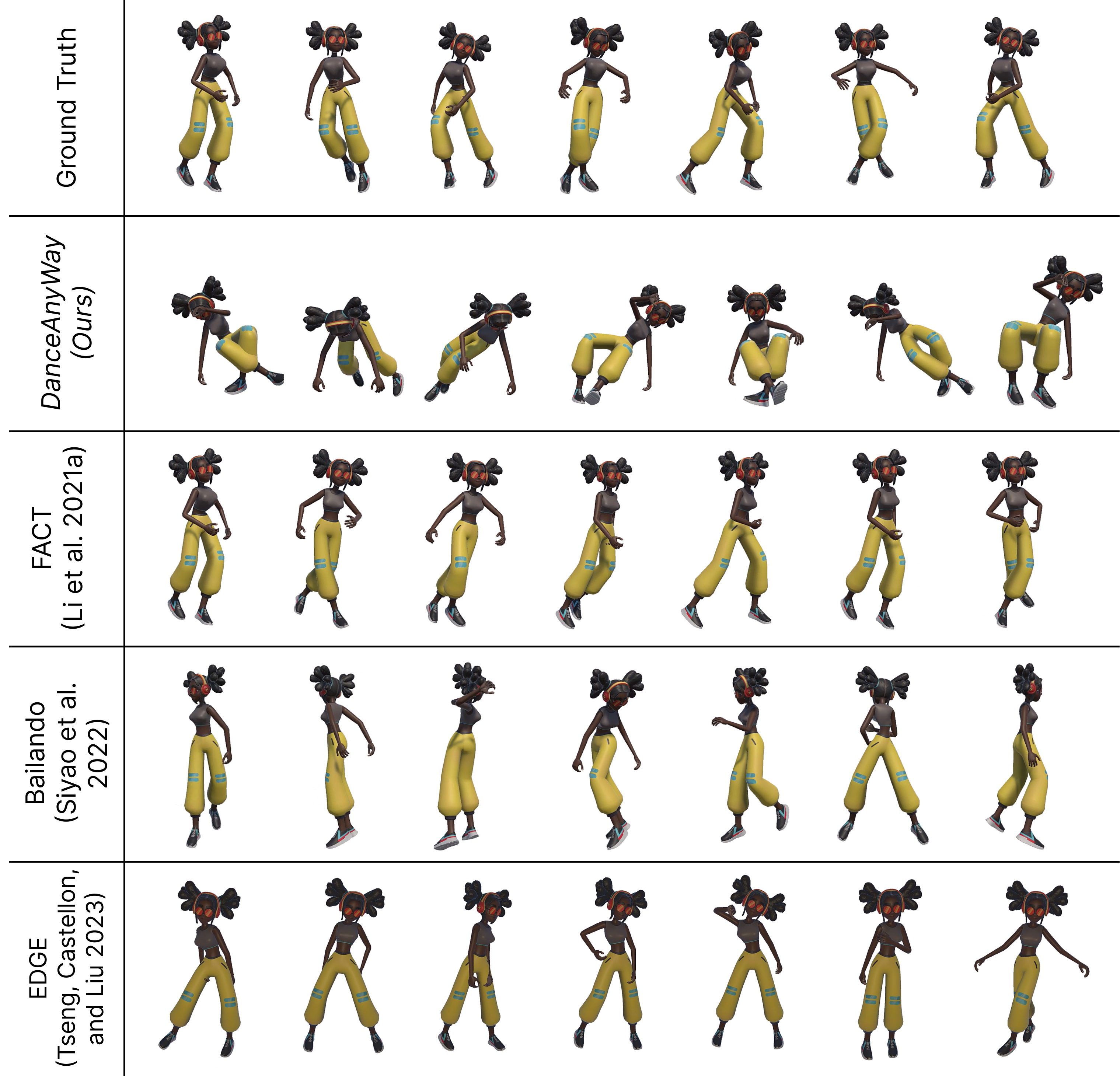}
    \caption{Visualizations on AIST++~\cite{aist++}. Sampled frames in a left-to-right sequence for one test sample. Our generated samples are better aligned with beats, more diverse, and have more plausible fine-grained details.}
    \label{fig:baseline_fig}
\end{figure}

\subsection{Inference}
During inference, we provide the input audio and the seed poses to our network. BPS generates the beat poses in one prediction step. RPS generates the entire dance and the root trajectories. To render our generated dances on human meshes, we follow the approach of~\cite{ai_choreo} to apply them on Mixamo characters (\textit{https://www.mixamo.com}).


\section{Experiments and Results}\label{sec:experiments}
We describe the benchmark dataset we evaluate on and our quantitative and qualitative performances.

\subsection{Dataset}
We use the benchmark AIST++ dataset~\cite{aist++}, a large-scale 3D dance dataset of paired music and pose sequences spanning ten dance genres. We use the official dataset splits for training and testing our model.

\subsection{Evaluation Metrics}
We use the following common evaluation metrics~\cite{ai_choreo,bailando,edge}.

\paragraph{Fr\'echet Inception Distance (FID).}
Following~\cite{ai_choreo,bailando}, we compute FID on both the kinetic ($k$) and the geometric ($g$) features to measure the generated motion quality relative to the ground truth.

\paragraph{Motion Diversity (MD).}
Following~\cite{ai_choreo,bailando,edge}, we compute MD on both the kinetic ($k$) and the geometric ($g$) features as well to measure the diversity of the generated dances relative to the ground truth.

\paragraph{Beat Alignment Score (BAS).}
BAS measures the temporal alignment of dances with the audio beats. It is essential to understand the rhythmic quality of the dances. We use the BAS implementation of prior work~\cite{ai_choreo,bailando,edge}.

\paragraph{Physical Foot Contact Score (PFC).}
We also report PFC, introduced by~\cite{edge} to measure the physical plausibility of the foot movements w.r.t. the ground plane by measuring foot sliding.

\begin{figure}[t]
    \centering
    \includegraphics[width=0.9\columnwidth]{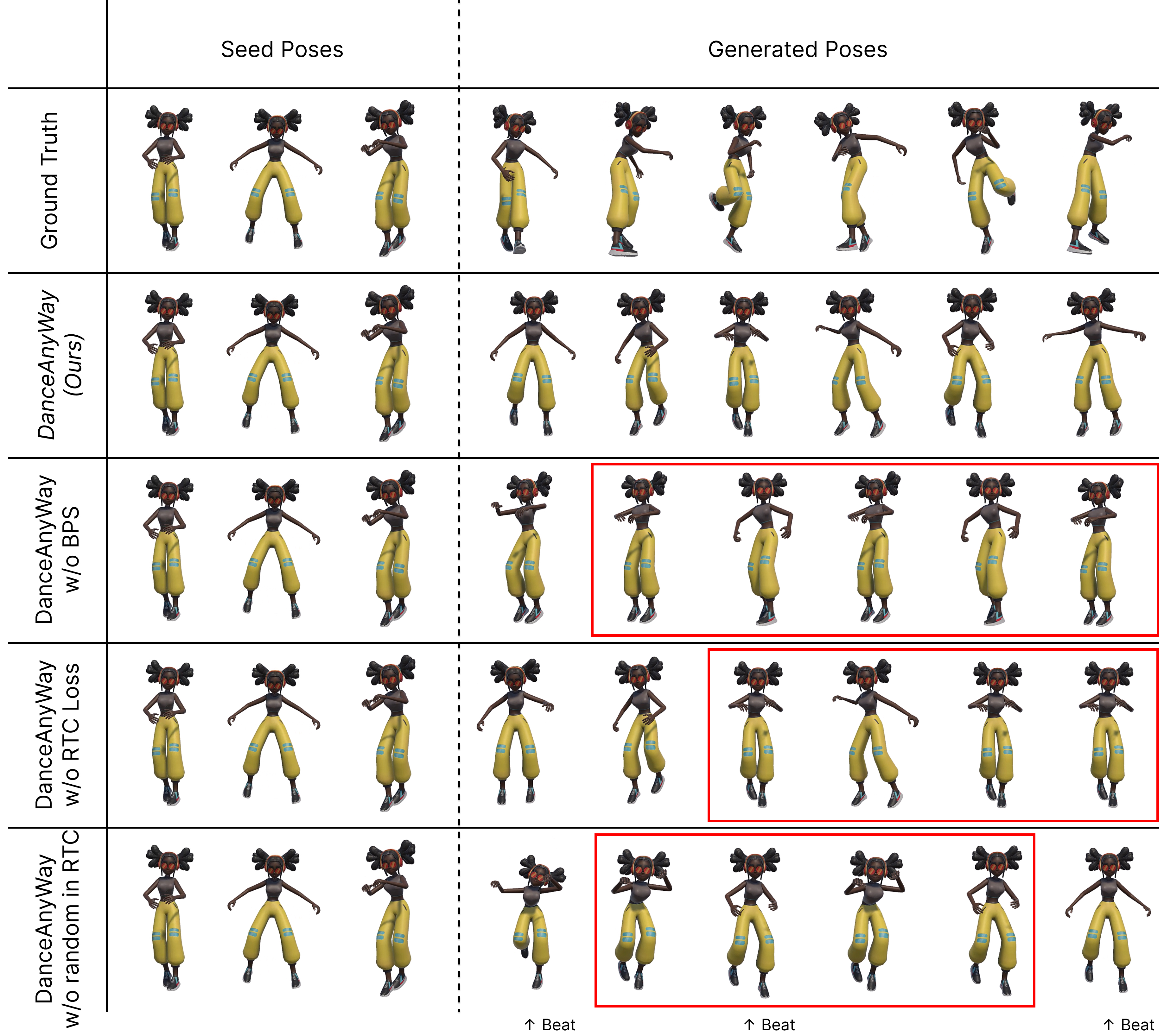}
    \caption{\methodname~Ablations on AIST++~\cite{aist++}. Sampled frames in a left-to-right sequence for one test sample. We highlight issues such as misalignment with beats (row 3), lack of motion diversity (row 4), and motion jitter (row 5) with red boxes.}
    \label{fig:ablation_fig}
\end{figure}

\subsection{Quantitative Evaluations}
We compare our proposed method, \methodname, with the baseline methods of Dance Transformers~\cite{transformer_dance1}, DanceNet~\cite{dancenet}, DanceRevolution~\cite{dancerev}, FACT~\cite{ai_choreo}, Bailando~\cite{bailando} and EDGE~\cite{edge}. We also evaluate three ablated versions of our method: without the BPS network, without the RTC loss, and assigning a fixed ``reference'' segment instead of using randomization for the RTC loss. We report all the results in Table~\ref{tab:quantitative_eval_aist}.

\begin{table}[t]
    \centering
    \begin{tabular}{lccccc}
        \toprule
        & GT & \textbf{\methodnameshort} & FACT & Bailando & EDGE \\
        \midrule
        Quality & 3.66 & \textbf{3.58} & 2.86 & 2.23 & \underline{3.36} \\
        Sync & 3.61 & \textbf{3.61} & 2.93 & 2.37 & \underline{3.60} \\
        \bottomrule
    \end{tabular}
    \caption{Perceptual Study Scores. Mean preferences on samples generated from AIST++~\cite{aist++} based on the quality of the dances and synchronization with the audio. Bold = best, underline = second-best among all methods.}
    \label{tab:perceptual_study_scores}
\end{table}

\paragraph{Comparison With Baselines.}
Compared to the best baseline of EDGE~\cite{edge}, our FID$_k$ and FID$_g$ scores are about $12\%$ and $7\%$ better respectively, our $\textrm{MD}_k$ and $\textrm{MD}_g$ scores are about $4\%$ and $1.5\%$ better respectively, and our BAS and PFS improve by about $22\%$ and $50\%$ respectively. We further demonstrate better beat alignment of our generated dances for a test sample in Fig.~\ref{fig:beat_align} and visualize snapshots from the generated sequences in Fig.~\ref{fig:baseline_fig}.

\paragraph{Comparison With Ablated Version Without BPS.}
In this ablation, we train only the RPS network with audio and seed poses to synthesize the dance sequences. Without BPS, the RPS network loses alignment with the audio beats as time progresses, leading to lower BAS (Table~\ref{tab:quantitative_eval_aist}, row 9). These results show the importance of BPS supplying the necessary beat information for long-term motion synthesis. 

\paragraph{Comparison With Ablated Version Without RTC Loss.}
In this ablation, we remove the RTC loss during training. We observe that the motion diversity drops rapidly as time progresses, and the network becomes susceptible to motion freezing and looping over limited motions after a few time steps, leading to lower MD (Table~\ref{tab:quantitative_eval_aist}, row 10). This also corroborates with how the network learns the RPS latent decoder space, consisting of feature sequences $\mathcal{Z}_{\mathcal{R}}$ (Eqn.~\ref{eq:RPS_latent_decoder}), with and without the RTC loss (Fig.~\ref{fig:rtc_tsne}). Using the RTC loss enables the network to avoid mode collapse and enforce diversity by clustering the features at different time steps within each sequence. Without the RTC loss, the network can still generate novel motions in sporadic bursts if it uses BPS. However, the overall motion diversity is limited, as we visualize on a random test sample in Fig.~\ref{fig:ablation_fig}, row 4.

\subsubsection{Comparison With Ablated Version Without Randomization of RTC Loss.}
In this ablation, we assign a fixed ``reference'' segment at the beginning of the sequence and compute the RTC loss w.r.t. this segment. The resultant synthesized dances are unstable as time progresses, changing the joint positions abruptly in an attempt to diversify from the reference segment. Going to the other extreme, minimizing the RTC loss across all segment pairs in each sequence leads to even higher temporal instability. To summarize, the lack of randomization leads to higher FID of the motions (Table~\ref{tab:quantitative_eval_aist}, row 11) and significant jitter (Fig.~\ref{fig:ablation_fig}, row 5).

\subsection{Perceptual Study}
We evaluate the perceived performance of our generated dances through a perceptual study with human participants. For each participant, we select eight random audios from the AIST++~\cite{aist++} test set and generate the corresponding dances using our method and the best-performing baselines methods of FACT~\cite{ai_choreo}, Bailando~\cite{bailando}, and EDGE~\cite{edge}. We show the participants these generated dances and the corresponding ground truth dances -- five dances in total for each audio -- in a random order unknown to them. We ask them to rate each dance for each audio on a five-point Likert scale on two aspects: motion quality and synchronization with the audio. To reduce inter-annotator variance, we also provide them guidelines on which aspects of the dances to focus on when assigning the Likert-scale scores. We detail these guidelines in our appendix.

\begin{figure}[t]
    \centering
    \includegraphics[width=\columnwidth]{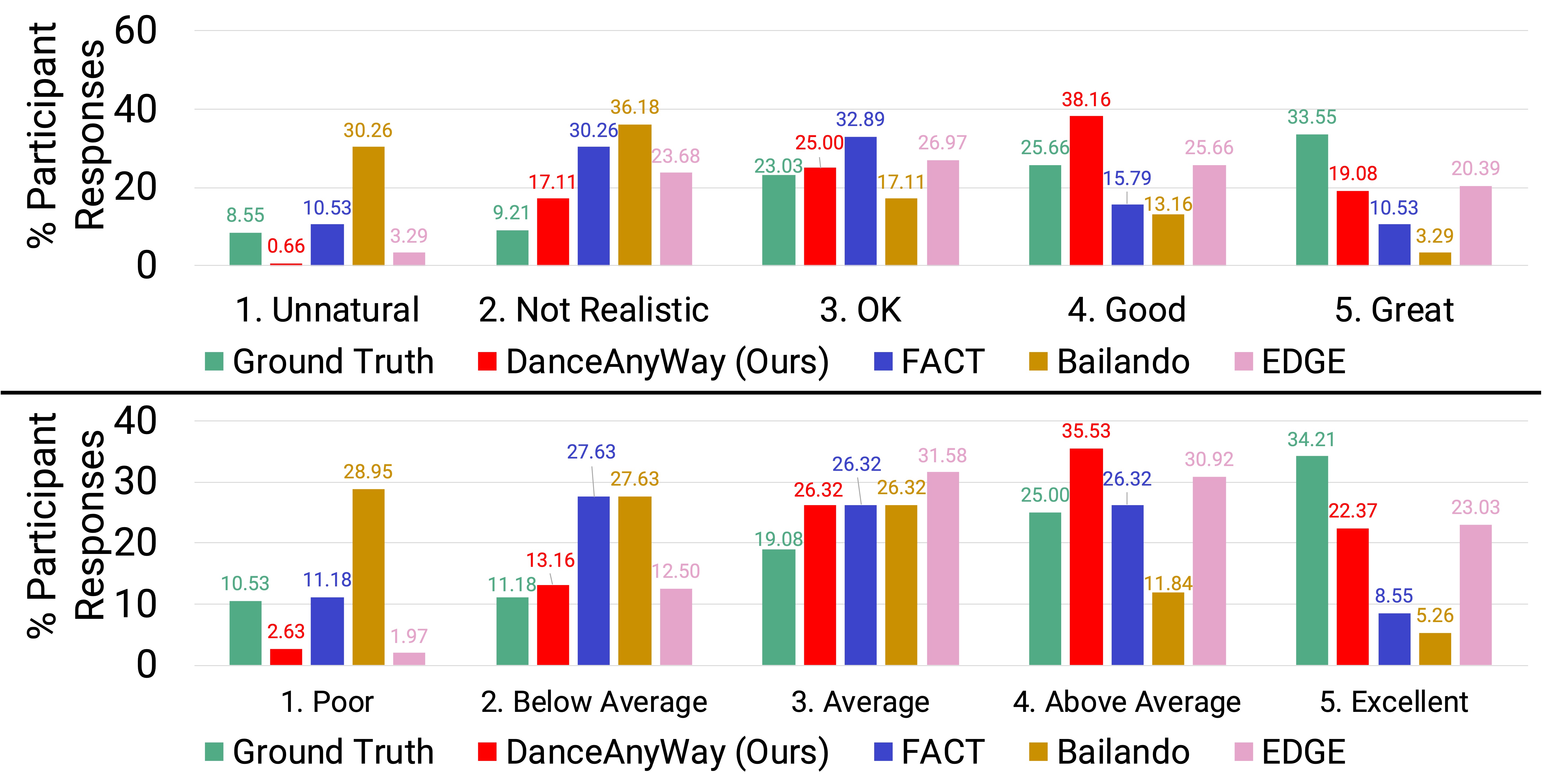}
    \caption{Perceptual Study Response Distributions. Distributions of the Likert-scale scores for all the methods and the ground truth on motion quality \textit{(top)} and synchronization with audio \textit{(bottom)}. We note EDGE and our method with the most responses of 3 or above among the methods.}
    \label{fig:perceptual_study_dists}
\end{figure}

We report results on $31$ responses to our perceptual study, discounting responses that failed our validation checks. $9$ identified as female, $20$ identified as male, $1$ identified as non-binary and $1$ did not disclose their gender. $16$ participants were between the ages of 18 and 24, $13$ between $25$ and $35$, and $2$ above $35$.
We report the mean Likert-scale scores for the methods and the ground truth in Table~\ref{tab:perceptual_study_scores}, and show the distribution of responses in Fig.~\ref{fig:perceptual_study_dists}. For motion quality, on average, participants preferred our generated dance motions $14\%$, $27\%$, and $4\%$ more compared to FACT, Bailando, and EDGE, respectively. They also marked our generated dances $3$ or more in motion quality for $82\%$ of the samples, compared to $59\%$ for FACT, $34\%$ for Bailando, and $73\%$ for EDGE. For synchronization, on average, participants preferred our generated dance motions $14\%$, $25\%$, and $0.2\%$ more compared to FACT, Bailando, and EDGE, respectively. They also marked our generated dances $3$ or more in synchronization for $84\%$ of the samples, compared to $61\%$ for FACT, $43\%$ for Bailando, and $85\%$ for EDGE.


\section{Conclusion and Future Work}
We have presented a novel learning method to synthesize beat-aligned, long-term 3D dances from audio. Through extensive quantitative and qualitative evaluations, we have demonstrated the state-of-the-art performance of our method on a benchmark dance dataset. In the future, we plan to extend our method to explicitly understand different dance styles and make the generation more controllable. We also plan to incorporate dancer-specific capabilities and human-human and human-object interactions.

\bibliography{danceanyway}

\appendix

\section*{Appendix}
We provide further details on our training loss functions and perceptual study setup.

\section{Training and Inference}
We detail the loss functions we use for training our network, the implementation details, and the testing procedure.

\subsection{Training Loss Functions}
We use pose motion and leg motion losses to first train our BPS, then use pose motion, leg motion, and generative adversarial losses to train our RPS, and finally, the root translation loss to train our translation decoder. We describe each of these losses below.

\paragraph{Pose Motion Loss.}\label{pose_motion_loss}
This is a linear combination of a distance metric $d_{pose}\parens{\cdot}$ between the ground truth and the generated 3D dance pose sequences and a distance metric $d_{vel}\parens{\cdot}$ between the corresponding temporal velocities. It captures the overall correctness and smoothness of the generated pose sequences in both spatial and temporal aspects. We write it as
\begin{equation}
    \mathcal{L}_{pm} = \lambda_{pose}d_{pose}\parens{\mathcal{U}, \hat{\mathcal{U}}} + \lambda_{vel}d_{vel}\parens{\Delta_f\mathcal{U}, \hat{\Delta_f\mathcal{U}}},
\end{equation}
where the $\Delta_f$ operator denotes forward differences between adjacent frames in the sequence, and $\lambda_{pose}$ and $\lambda_{vel}$ are experimentally determined fixed weights. For BPS, we have $\mathcal{U} = \mathcal{U}_{-B_S}$ and $\hat{\mathcal{U}} = \hat{\mathcal{U}}_{-B_S}$ and set both $d_{pose}$ and $d_{vel}$ to be the mean squared error (MSE). For RPS, we have $\mathcal{U} = \mathcal{U}$ and $\hat{\mathcal{U}} = \hat{\mathcal{U}}$ and set both $d_{pose}$ and $d_{vel}$ to be the smooth $\ell_1$ loss. For both BPS and RPS, we use $\lambda_{pose} = \lambda_{vel} = 1$.

\paragraph{Leg Motion Loss.}\label{leg_motion_loss}
This is a linear combination of the position and velocity losses operating exclusively on the angles $\theta_{fs}$ and angular velocities $\omega_{fs}$ between the femur and the shin bones of both legs. It provides additional constraints on leg movements and reduces foot sliding artifacts. We write it as
\begin{equation}
    \mathcal{L}_{lm} = \lambda_{pose}d_{pose}\parens{\theta_{fs}, \hat{\theta_{fs}}} + \lambda_{vel}d_{vel}\parens{\omega_{fs}, \hat{\omega_{fs}}},
\end{equation}
where we use the smooth $\ell_1$ loss for both $d_{pose}$ and $d_{vel}$ in both BPS and RPS, and set $\lambda_{pose} = 0.3$ and $\lambda_{vel} = 0.7$.

\paragraph{Generative Adversarial Losses.}
We use the conventional generator loss $\mathcal{L}_{gen}$ and discriminator loss $\mathcal{L}_{disc}$ to train the generator and discriminator, respectively, in our RPS. For completeness, we write them as
\begin{align}
    \mathcal{L}_{gen} &= -\mathbb{E}\bracks{\log\parens{\mathrm{Disc}\parens{\hat{\mathcal{U}}}}}, \\
    \mathcal{L}_{disc} &= -\mathbb{E}\bracks{\log\parens{\mathrm{Disc}\parens{\mathcal{U}}}} - \mathbb{E}\bracks{\log\parens{1-\mathrm{Disc}\parens{\hat{\mathcal{U}}}}},
\end{align}
where $\mathrm{Disc}\parens{\cdot}$ denotes our entire discriminator network.

\paragraph{Root Translation Loss.}\label{root_translation_loss}
This is a linear combination of the position and velocity losses operating exclusively on the sequences of root positions in the global coordinates. It ensures the correctness and smoothness of the trajectories generated by our translation decoder. We write it as
\begin{equation}
    \mathcal{L}_{rt} = \lambda_{pose}d_{pose}\parens{\mathcal{R}, \hat{\mathcal{R}}} + \lambda_{vel}d_{vel}\parens{\Delta_f \mathcal{R}, \Delta_f \hat{\mathcal{R}}},
\end{equation}
where we use MSE for $d_{pose}$ and the smooth $\ell_1$ loss for $d_{vel}$, and set $\lambda_{pose} = \lambda_{vel} = 1$.

\paragraph{Total Training Loss for BPS and RPS Generator.}
For our BPS, we use the total loss $\mathcal{L}_{BPS}$ as
\begin{equation}
    \mathcal{L}_{BPS} = \lambda_{pm}\mathcal{L}_pm + \lambda_{lm}\mathcal{L}_{lm},
\end{equation}
and for our RPS generator, we use the total loss $\mathcal{L}_{RPSGen}$ as
\begin{equation}
    \mathcal{L}_{RPSGen} = \lambda_{pm}\mathcal{L}_pm + \lambda_{lm}\mathcal{L}_{lm} + \lambda_{gen}\mathcal{L}_{gen},
\end{equation}
where we experimentally set the weights $\lambda_{pm}=5$, $\lambda_{lm}=3e-3$, and $\lambda_{gen}=5e-2$.

\section{Perceptual Study Setup}
We report the set of guidelines we provide our participants for the perceptual study in Table~\ref{tab:study_qs}. These guidelines initiate
the participants on which aspects of the dances to focus on, thus bounding the variances in their responses and reducing inter-annotator disagreement.

\begin{table*}[t]
    \centering
    \begin{tabular}{rL{0.3cm}lL{4.5cm}L{0.3cm}lL{4.5cm}}
        \toprule
        Score && \multicolumn{2}{c}{Motion Quality} && \multicolumn{2}{c}{Synchronization} \\
        \cmidrule{1-1}
        \cmidrule{3-4}
        \cmidrule{6-7}
        1 && Very Unnatural & Broken dance motions, Broken arms or legs, torso at impossible angles, etc. && Poor & No or arbitrary movements \\
        2 && Not Realistic & Unnatural movements, freezing of dance motions, self-collisions, mismatched proportions, etc. && Below Average & Most of the movements appear to be out-of-sync from the audio \\
        3 && Looks OK & No serious problems, but the
        motion does not look very appealing && Average & Some movements are in sync, and some movements are out-of-sync \\
        4 && Looks good & No problems and the dance motions look natural && Above Average & Most of the movements appear to be in sync with the audio \\
        5 && Looks great! & The dance motions look like they could be from a real person && Excellent! & Movements appear to be well-synced with the audio \\
        \bottomrule
    \end{tabular}
    \caption{Perceptual Study Guidelines. We report the guidelines for the 5-point Likert scale scores for both the aspects of motion quality and synchronization.}
    \label{tab:study_qs}
\end{table*}

\end{document}

%% file: macros_danceanyway.tex
\usepackage{amsfonts}
\usepackage{amsmath}
\usepackage{amsthm}
\usepackage{array}
\usepackage{bm}
\usepackage{booktabs}
\usepackage{color}
\usepackage{enumitem}
\usepackage{float}
\usepackage{mathtools}
\usepackage{multirow}
\usepackage{subcaption}
\usepackage[dvipsnames]{xcolor}
\usepackage{xurl}

\theoremstyle{definition}

\newcolumntype{L}[1]{>{\raggedright\let\newline\\\arraybackslash\hspace{0pt}}m{#1}}
\newcolumntype{C}[1]{>{\centering\let\newline\\\arraybackslash\hspace{0pt}}m{#1}}
\newcolumntype{R}[1]{>{\raggedleft\let\newline\\\arraybackslash\hspace{0pt}}m{#1}}

\setlist[itemize]{noitemsep, topsep=0pt}
\setlist[enumerate]{noitemsep, topsep=0pt}

\newcommand{\sota}{state-of-the-art}
\newcommand{\methodname}{DanceAnyWay}
\newcommand{\methodnameshort}{DAW}

\newcommand{\parens}[1]{\left(#1\right)}
\newcommand{\braces}[1]{\left\{#1\right\}}
\newcommand{\bracks}[1]{\left[#1\right]}
\newcommand{\modulus}[1]{\left\vert#1\right\vert}

\newcommand{\ceil}[1]{\left\lceil#1\right\rceil}


%% file: abstract.tex
\begin{abstract}
    We present \methodname, a generative learning method to synthesize beat-guided dances of 3D human characters synchronized with music. Our method learns to disentangle the dance movements at the beat frames from the dance movements at all the remaining frames by operating at two hierarchical levels. At the coarser ``beat'' level, it encodes the rhythm, pitch, and melody information of the input music via dedicated feature representations only at the beat frames. It leverages them to synthesize the beat poses of the target dances using a sequence-to-sequence learning framework. At the finer ``repletion'' level, our method encodes similar rhythm, pitch, and melody information from all the frames of the input music via dedicated feature representations. It generates the full dance sequences by combining the synthesized beat and repletion poses and enforcing plausibility through an adversarial learning framework. Our training paradigm also enforces fine-grained diversity in the synthesized dances through a randomized temporal contrastive loss, which ensures different segments of the dance sequences have different movements and avoids motion freezing or collapsing to repetitive movements. We evaluate the performance of our approach through extensive experiments on the benchmark AIST++ dataset and observe improvements of about $7\%-12\%$ in motion quality metrics and $1.5\%-4\%$ in motion diversity metrics over the current baselines, respectively. We also conducted a user study to evaluate the visual quality of our synthesized dances. We note that, on average, the samples generated by our method were about $9-48\%$ more preferred by the participants and had a $4-27\%$ better five-point Likert-scale score over the best available current baseline in terms of motion quality and synchronization. Our source code and project page are available at https://github.com/aneeshbhattacharya/DanceAnyWay.
\end{abstract}